\documentclass{SCIS2024}

\usepackage[utf8]{inputenc}
\usepackage[
    backend=biber,
    style=numeric,
    maxnames=3,   
    minnames=3,   
    giveninits=false, 
]{biblatex}

\DeclareFieldFormat{labelnumberwidth}{#1\hspace{\labelsep}} 

\defbibenvironment{bibliography}
  {\list{\printtext[labelnumberwidth]{\printfield{labelnumber}}}
     {\setlength{\labelwidth}{\labelnumberwidth}
      \setlength{\leftmargin}{\dimexpr\labelwidth+\labelsep\relax}%
      \advance\leftmargin\labelsep
      \setlength{\itemsep}{0pt}
      \setlength{\parsep}{0pt}
      \setlength{\parskip}{0pt}
      \setlength{\topsep}{0pt}
      \footnotesize\sloppy}} 
  {\endlist}
  {\item\relax}

\defbibheading{bibliography}[\refname]{%
  \section*{\small #1} 
}

\DeclareFieldFormat[article,inbook,incollection,inproceedings,patent,thesis,unpublished]{title}{#1}
\DeclareFieldFormat{pages}{#1}
\DeclareCiteCommand{\cite}
  {\usebibmacro{prenote}}
  {\bibopenbracket\printfield{labelnumber}\bibclosebracket} 
  {\multicitedelim}
  {\usebibmacro{postnote}}
\DeclareFieldFormat{labelnumberwidth}{#1} 

\DeclareBibliographyDriver{article}{%
  \printnames{author} 
  \newunit \newblock
  \printfield{title} 
  \newunit \newblock
  \printfield{journal} 
  \newunit \newblock
  \printfield{year} 
  \newunit \newblock
  \printfield{volume} 
  \printfield{number} 
  \setunit{:}\newblock
  \printfield{pages} 
  \finentry}

\DeclareFieldFormat[inproceedings]{booktitle}{#1} 
\DeclareBibliographyDriver{inproceedings}{%
  \printnames{author} 
  \newunit \newblock
  \printfield{title} 
  \newunit \newblock
  \printtext{In:} 
  \printfield{booktitle} 
  \setunit{\addcomma\space} 
  \printfield{location} 
  \setunit{\addcomma\space} 
  \printfield{year} 
  \setunit{\adddot\space} 
  \printfield{pages} 
  \finentry}

\usepackage{tcolorbox}
\usepackage{multirow}
\tcbuselibrary{skins, breakable, theorems}

\newcommand{\lqy}[1]{{#1}}

\begin{document}
\ArticleType{RESEARCH PAPER}
\Year{2024}
\Month{}
\Vol{}
\No{}
\DOI{}
\ArtNo{}
\ReceiveDate{}
\ReviseDate{}
\AcceptDate{}
\OnlineDate{}

\title{CupCleaner: A Hybrid Data Cleaning Approach for Comment Updating}{CupCleaner:  A Hybrid Data Cleaning Approach for Comment Updating}

\author[1]{Qingyuan Liang}{}
\author[2]{Zeyu Sun}{{zeyu.zys@gmail.com}}
\author[1]{Qihao Zhu}{}
\author[1]{Junhao Hu}{}
\author[1]{Yifan Zhao}{}
\author[1]{Yakun Zhang}{}
\author[1]{\\Lu Zhang}{{zhanglucs@pku.edu.cn.}}

\AuthorMark{Qingyuan Liang}

\AuthorCitation{Qingyuan Liang, Zeyu Sun, Qihao Zhu, et al}


\address[1]{Key Laboratory of High Confidence Software Technologies (Peking University), Ministry of Education; \\ School of Computer Science, Peking University, Beijing {\rm 100871}, China}
\address[2]{National Key Laboratory of Space Integrated Information System, Institute of Software, \\Chinese Academy of Sciences, Beijing {\rm 100190}, China}

\abstract{

\lqy{Comment updating is an emerging task in software evolution that aims to automatically revise source code comments in accordance with code changes. This task plays a vital role in maintaining code-comment consistency throughout software development. Recently, deep learning-based approaches have shown great potential in addressing comment updating by learning complex patterns between code edits and corresponding comment modifications.
However, the effectiveness of these learning-based approaches heavily depends on the quality of training data. Existing datasets are typically constructed by mining version histories from open-source repositories such as GitHub, where there is often a lack of quality control over comment edits. As a result, these datasets may contain noisy or inconsistent samples that hinder model learning and generalization.}
In this paper, we focus on cleaning existing comment updating datasets, considering both the data's characteristics in the updating scenario and their implications on the model training process. 
We propose a hybrid statistical approach named CupCleaner (Comment UPdating's CLEANER) to achieve this purpose. 
Specifically, we combine static semantic information within data samples and dynamic loss information during the training process to clean the dataset. 
Experimental results demonstrate that, on the same test set, both the individual static strategy and the dynamic strategy can significantly filter out a portion of the data and enhance the performance of the model. 
Furthermore, employing a model ensemble approach can combine the characteristics of static and dynamic cleaning, further enhancing the performance of the model and the reliability of its output results. 

}


\maketitle

\section{Introduction}

Software evolution is a pivotal and indispensable process in the field of software engineering~(SE), as it encompasses the continuous development and improvement of software systems~\cite{lientz1978softwaremainten, bennett2000softwareevolution}. Throughout the process of software evolution, a substantial volume of data is generated and recorded, holding substantial research value. 
These data can be leveraged to support various data-driven research tasks related to software evolution, such as automated code repair and code review~\cite{tufano2021codereview, codereview1, just2014defects4j, recoder, divot5}. 
Among them, comment updating has been recognized as an important and emerging research task during software development and maintenance~\cite{acl20, aaai21}. 
This task aims to automatically generate new code comments based on existing code comments and code changes. On one hand, automatic comment updating can save time and effort for programmers to manually write new comments. Unlike directly generating comments, comment updating can take into account more information, such as code changes and old comments. Thus, this task is more practical and relevant to the daily development scenarios of programmers. On the other hand, if comments are not updated in a timely manner, they may mislead future development activities. For example, if obsolete TODO comments are not removed, they may introduce new bugs in future development~\cite{todocomm}.

\begin{figure}[t]
\centering
\centerline{\includegraphics[scale=0.9]{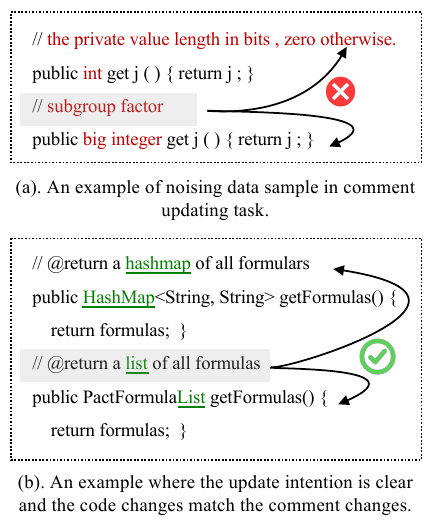}}
\caption{Two data examples in the comment updating task.}
\label{fig_example}
\end{figure}

Currently, with the advancement of artificial intelligence~(AI) technology~(e.g., deep learning), particularly in the area of code pre-training, there is a promising potential to offer an effective solution to the comment updating task.
High-quality data is a key factor in deep learning-based approaches \cite{aidata, scis_deeplearning, wang2024advanced}, and higher quality data often leads to more powerful and more reliable models, while lower quality data can have the opposite effect.
Data collection during the software evolution process presents challenges, as handling different versions of code and comments often makes it difficult to ensure the quality of data.
Datasets related to the comment updating task are typically crawled from online repositories, such as GitHub, by extracting commit history versions. This also makes it easier for noisy data to be included in datasets for the comment updating task~\cite{conala, githubnoise1,githubnoise2, githubnoise3}. 
Figure~\ref{fig_example} illustrates two data samples in the comment updating task. In subfigure~(a), the sample is considered noisy, as the comment fails to reflect the functionality of the source code (e.g., compared to the bottom code, ``subgroup factor'' comment is unrelated to its functionality). Additionally, the comment change does not capture the variations between code changes~(e.g., compared to the original comment, the changes in the updated comment fail to reflect the transition in the code from ``int'' to ``big integer''). 
Subfigure~(b) presents a clear sample that code changes match the comment changes, where the change from ``HashMap'' to ``List'' is reflected in both the comments and the code.
If the dataset contains a large amount of noise like that shown in subfigure~(a), the results generated by a model trained with the dataset would also be misleading. 
Therefore, to enhance the reliability of the models, it is crucial to improve the quality of the dataset by cleaning datasets.


\lqy{
While recent studies have explored data cleaning for code-related tasks, such as comment generation or code search~\cite{codesearch, commclean, codecomment}, these approaches primarily rely on heuristic rules targeting individual code-comment pairs. Although effective to some extent in identifying clearly low-quality samples, such rule-based methods face fundamental limitations when applied to the comment updating task.
First, these methods often fail to capture the semantic alignment between code changes and comment changes. Unlike static code-comment matching tasks, comment updating inherently involves reasoning over differences between old and new code, and the corresponding changes in comments. Simple surface-level rules~(e.g., keyword matching or edit distance thresholds) are insufficient for identifying whether a comment change is meaningful or consistent with the associated code edit. This highlights the need for cleaning strategies that account for semantic-level relationships between code and comment evolution.
Second, existing approaches overlook the rich signals available during model training. In practice, low-quality data often reveals itself through training dynamics, for example, by producing higher loss values, slower convergence, or degraded generalization. However, rule-based methods operate independently of model feedback, treating all training samples uniformly regardless of how the model learns from them. This fails to leverage the opportunity to evaluate data quality in a model-aware manner.
}

\begin{figure*}[t]
\centering
\centerline{\includegraphics[scale=0.14]{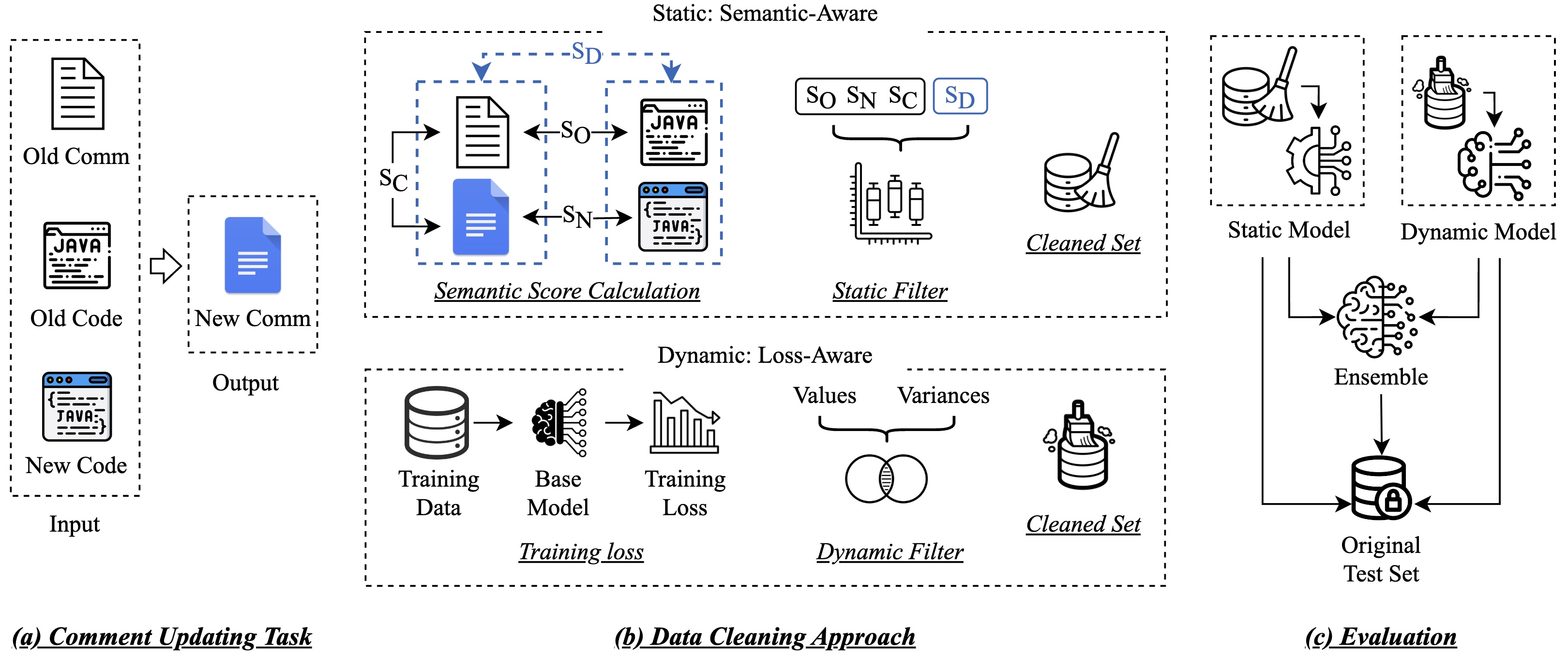}}
\caption{Overview of our research procedure.}
\label{fig_overview}
\end{figure*}

In this paper, we propose CupCleaner~(\textbf{C}omment \textbf{UP}dating’s \textbf{CLEANER}), a hybrid data cleaning approach that leverages both static and dynamic information of the data. Specifically, we take the semantics of the data as the static information~(i.e., semantic-aware), while in terms of dynamic information, we consider the loss of the model during the training process~(i.e., loss-aware).
\textbf{The static cleaning strategy aims to address the first limitation, while the dynamic cleaning strategy aims to tackle the second limitation.}
By incorporating these two aspects, CupCleaner aims to enhance the data cleaning process effectively.
Subfigure~(a) in Figure~\ref{fig_overview} depicts the task of comment updating, and Subfigure~(b) shows an overview of the CupCleaner.
During the process of leveraging semantic information for cleaning, we focus on two main types of noise in comment updating: weak correlation within comments or code-comment pairs~(i.e., black solid line), and weak correlation between code changes and comment changes~(i.e., blue dashed line). Then we design a heuristic rule based on semantic information to compute a score for each data sample in the comment updating dataset, and filter out the ones with low scores. 
During the data cleaning process using model training loss, we initially train a base model to derive the loss and its variations for each data sample throughout the training process. Subsequently, we analyze by considering both the values and variances of the loss, enabling us to filter out data samples that pose challenges for the model to learn effectively. 
As depicted in Subfigure~(c) of Figure~\ref{fig_overview}, CupCleaner employs static and dynamic statistical information to filter the comment updating dataset. 
We train models using datasets that are cleaned separately using static and dynamic information, and conduct testing on the original test set to evaluate their performance.
Additionally, we further integrate the two cleaning approaches at the model level to enhance the robustness and performance of the final model.
It is worth noting that CupCleaner is an unsupervised approach for improving data quality, which allows it to work without the need for pre-prepared or assumed high-quality data.

To evaluate CupCleaner, we first conduct a human evaluation by mixing the noise data and high-quality data. The evaluation results show that, with a maximum score of 5, the average score for the noise data identified by CupCleaner is only 2.5, while the high-quality data received a score of 4.3.
Then we conduct experiments on two comment updating datasets using two representative models: PLBART~\cite{plbart} and CodeT5~\cite{codet5}. 
The experimental results demonstrate that combining static and dynamic cleaning approaches effectively enhances the model's performance. 
For example, on the dataset from \textit{Panthaplackel et al., 2021}, compared to models trained on original data, the PLBART model trained on data cleaned by CupCleaner increases the exact match metric from 33.33\% to 38\%. Similarly, the CodeT5 model trained on cleaned data also raises the exact match metric from 41.33\% to 45.33\%, surpassing the performance of the dedicated code editing pre-trained model~(i.e., CoditT5~\cite{coditt5}).
In addition, we separately evaluate the impact of static and dynamic cleaning strategies on model training. Experimental results indicate that both static and dynamic cleaning strategies enhance model performance. Furthermore, combining these two cleaning strategies further enhances the model's performance and generalization capabilities.
Besides, we also explore the time consumption of cleaning data using CupCleaner. Comparative results indicate that the time spent cleaning a dataset is acceptable compared to fine-tuning models on the dataset.
In summary, training with data cleaned using CupCleaner can provide higher-quality data within a limited time frame. This enables the same model to exhibit improved performance, thus enhancing the model's reliability.

We summarize our contributions as follows:
\begin{itemize}
    \item \textbf{Data Cleaning}: We propose semantic-aware static data cleaning strategies and loss-aware dynamic data cleaning strategies for data cleaning. To the best of our knowledge, we are the \textbf{first} to develop a data cleaning approach that takes into account both the characteristics of the data and its impact on the model training process.
    \item \textbf{Model Ensemble}: We ensemble different cleaning strategies at the model level, selecting the superior output from models trained with two different cleaning strategies as the final output of the model.
    \item \textbf{Evaluation}: We conduct experiments on two different comment updating datasets. Human evaluation results suggest that the scores of the filtered-out data are significantly lower compared to those of the retained data. The experimental results indicate that both cleaning strategies improve data quality, and combining the two strategies further enhances the model's performance and generalization ability.
    \item The data and code are available at~\cite{cupcleaner}.
\end{itemize}

\section{Related Work}
Recently, an increasing number of researchers have become aware of the beneficial impact of high-quality data on model training~\cite{guo2024deepseekcoder,grammarcoder,liang2024condor}.
Meanwhile, an increasing number of AI for SE studies focus on editing or updating related tasks. 
Below, we discuss the two most related works to our paper: code data cleaning techniques and updating-related tasks.

\subsection{Code Data Cleaning}
Data quality is crucial for deep learning models, as only high-quality datasets can comprehensively exploit the potential of deep learning techniques.
In recent years, many researchers in the field of AI for SE focused on how to improve the quality of code-related data.
\textit{Sun et al.} \cite{codesearch} conduct an empirical study on the dataset for code search and find that more than one-third of the queries contain noise. To improve the quality of code search datasets, they propose a data cleaning framework by combining syntactic and semantic filters.
The experimental results show that the performance of code search is significantly improved with the cleaned dataset. \textit{Shi et al.} \cite{codesummary} focus on evaluation and data issues in code summarization tasks, and find that the pre-processing choices of the data have a significant impact on the experimental results. 
\textit{Huang et al.}\cite{codecomment} analyze the issue of whether code needs to be commented or not.
They find that the percentage of methods with both header and internal comments in software systems is low.
Then they design an approach to determining whether code needs to be commented by considering structural features, syntactic features, and textual features.
\textit{Shi et al.} \cite{commclean} subsequently analyze the noisy data in code comments and improve the quality of code comments.
They propose an automatic code-comment cleaning tool and conduct experiments on four code summarization datasets.
\textit{Xu et al.}~\cite{ocddata_quality} utilize character-level differences in comment changes to clean the data used for the obsolete comment detection task. 
Specifically, character-level differences can be used to filter out minor comment changes. These filtered comment changes serve as non-obsolete comment samples and are used to train a classifier for determining whether comments are obsolete.

However, applying these approaches to updating-related tasks may primarily face the following two limitations. First, the approaches focusing on the correlation between individual code and comments may not align well with the updating-related scenario, as the comment updating task involves changes in both code and comments. Second, relying solely on heuristic rules or isolated semantics may not adequately capture the overall data quality. It is also crucial to consider the impact of data samples on the model training process.
Different from these approaches that focus on cleaning individual comments or code, our proposed CupCleaner combines semantic-based static information and model loss-based dynamic information to consider the quality of data comprehensively. 
Additionally, CupCleaner is a purely unsupervised approach that does not require extra high-quality data to train a data quality discriminator. This allows it to be more flexibly adapted to various scenarios, such as being used to provide higher-quality datasets before \textit{Xu et al.}~\cite{ocddata_quality}'s work.

\subsection{Comment Updating Tasks}
\lqy{Updates are a fundamental part of software development, and many intelligent software engineering tasks revolve around automatically generating or modifying artifacts during the evolution process. For example, automatic program repair focuses on updating buggy code to restore correctness~\cite{repair1,repair2,repair3,repair4,repair5,recoder}, while automated code review assists developers in updating code based on review suggestions~\cite{codereview1,codereivew2,codereviewer}.}

\lqy{Among these, comment updating has recently emerged as a focused task that aims to automatically revise natural language comments in response to changes in the corresponding source code~\cite{ase20,acl20,aaai21,coditt5}.  One of the earliest efforts in this direction is the work by \textit{Panthaplackel et al.}~\cite{acl20}, which formulates the task as generating an updated comment based on the code changes and its associated old comment. To support this task, the authors constructed a large-scale dataset by mining synchronized comment-code edits from commit histories of open-source Java projects. This dataset focuses primarily on \texttt{@return} comments and provides aligned~(old comment, old code, new code, new comment) tuples for supervised learning.
Subsequent work by the same group~\cite{aaai21} shifts focus to the problem of detecting whether a comment is consistent with its corresponding code. Although the primary task is framed as binary classification~(i.e., consistent vs. inconsistent), the dataset construction process involves identifying cases where code has changed but comments remain outdated, making it possible to extract a large number of comment update samples. Compared to~\cite{acl20}, this dataset expands the coverage beyond \texttt{@return} to include additional comment types such as \texttt{@param} and summary comments. CoditT5~\cite{coditt5} later reused this dataset by selecting samples where both code and comments were modified, repurposing them for the comment updating task and training a sequence-to-sequence model for generating revised comments.}

\lqy{However, datasets supporting such tasks are often mined from the commit versions of open-source repositories. Due to the complexity of version iterations and the diversity of update intentions, such datasets are more prone to incorporating noisy data\cite{conala, githubnoise1,githubnoise2, githubnoise3}. 
To evaluate and enhance the quality of the comment updating task, we propose CupCleaner for data cleaning for comment updating.
Unlike studies that focus on designing specialized deep learning approaches to enhance model generation capabilities, this paper aims to clean existing datasets to improve the performance and reliability of the original models.
Our experimental results indicate that CupCleaner can effectively filter out noise from the comment updating datasets, and models trained on the cleaned data exhibit better performance.}

\section{Approach}
To measure data quality and clean comment updating datasets, we consider the real requirements in comment updating and design the CupCleaner to automate the data cleaning process. The overall framework of our approach is shown in Subfigure~(b) in Figure~\ref{fig_overview}. 
Our data cleaning process consists of two strategies:
The static data cleaning strategy, which utilizes semantic-based heuristic rules to filter the data, and the dynamic data cleaning strategy, which leverages loss information from the model training process to filter the data.

\subsection{Static Strategy}
In the static strategy, we design a semantic-aware criterion to calculate the quality scores for each data sample.

\subsubsection{Semantic Score Calculation}
Different from the data cleaning process that solely caters to individual comment-code pairs, samples of the comment updating task require considering the correlation not only between code and comments but also between code changes and comment changes. 
Therefore, we categorize the data noise into two types. 
The first type of noise exists between comments or comment-code pairs within each data sample. We evaluate their correlation by calculating their internal semantic similarity.
The second type of weak correlation exists between comment changes and code changes, and we evaluate this correlation based on \textit{diff} information. Below, we discuss the details of how to calculate the score and build filters.

\paragraph{Noise existing between comments or comment-code pairs (Type I)}
This type of noise arises from common issues with data quality, such as invalid or incorrect comments. We leverage the inherent semantic similarity within them to identify and address this type of noise.
To determine whether there is invalid data within the comments or code, we use the code pre-trained models to convert them into embedding vectors to represent semantics.
Currently, there are many pre-trained models used to represent the semantics of texts~\cite{bert, roberta, codebert, gcb, unixcoder}. We choose the state-of-the-art model pre-trained on code-related datasets,  UniXcoder~\cite{unixcoder}, to calculate the semantics of comments or code. We use the cosine similarity of semantic embeddings to describe the correlation between old and new comments, and between comments and code, respectively. 
Below, we present the specific calculation details.
\begin{align}
& S^i_{C} = Cos(M(comm^i_{old}),M(comm^i_{new})) \\
& S^i_{O} = Cos(M(comm^i_{old}),M(code^i_{old})) \\
& S^i_{N} = Cos(M(comm^i_{new}),M(code^i_{new})) \\
& S^i_{1} = Average(S^i_{C}, S^i_{O}, S^i_{N}) \label{score_1}
\end{align}
Where $comm^i$ represents the i-th comment, $code^i$ represents the i-th snippet of code, $M$ represents the pre-trained model~(i.e., UniXcoder), and $Cos$ represents cosine similarity. 
We compute the similarity $S^i_{C}$ between comments, the similarity $S^i_{O}$ between the old comments and old code, and the similarity $S^i_{N}$ between the new comments and new code. The average of these three similarities is used as the score $S_1$ for Type 1 noise.

\paragraph{Noise existing between comment changes and code changes (Type II)}
This type of noise is unique to the comment update task and is mainly caused by inconsistencies between comment changes and code changes. We use the word-level \textit{diff} information as the basic information for calculating the semantics, and use it to evaluate the data quality score. We follow the approach proposed by \cite{acl20} to get the basic \textit{diff} information between comments and code. 
For the semantics of \textit{diff} information, we concatenate the changed words in the \textit{diff} into a string and use UniXcoder to calculate the semantics embeddings.
We utilize the following formula to calculate the score for the semantics of \textit{diff} information.
\begin{align}
& S^i_2 = S^i_D = Cos(M(d_{comm}^i),M(d_{code}^i))
\end{align}
Where $d_{comm}$ denotes the \textit{diff} information among comments in the $i$-th data sample, while $d_{code}^i$ denotes the \textit{diff} information among source code in the $i$-th data sample.

\paragraph{Final score}
We consider both types of noise mentioned above and calculate a unified score to assess the data quality of comment updating. Specifically, we calculate two scores and determine the maximum score among them as the final score for each data sample. The formal representation is as follows:
\begin{align}
& S^i= Max(S^i_1, S^i_2) \label{final_score}
\end{align}
Where the $S^i$ represents the final score of the $i$-th data sample.
We use $Max$ to represent that the semantics of the \textit{diff} only needs to be combined with the maximum value of the semantic similarity between comments and code.
This is actually intuitive, as once the semantics of changes is similar, or if there is a high similarity between the comments or code, then it is highly likely to be a data sample that fits the update scenario.

\subsubsection{Static Filter}

In our semantic-aware approach, we quantify the internal correlations of data by evaluating the score of each data sample. Building upon this, we adopt a specific filtering mechanism to ensure that only data samples with higher scores are retained, while those with lower scores are excluded. To be specific, we choose the lower quartile~($Q_1$) as a reference point and filter out data below this threshold by a certain margin. The formal computation is expressed as follows: 
$$X_{low} = Q_1 - k \times \text{IQR}$$, 
where $X_{low}$ denotes the final threshold and $\text{IQR}$ represents the interquartile range~(i.e., the gap between the upper quartile and the lower quartile). 
The parameter $k$ ($k > 0$) acts as a coefficient that controls the strictness of data filtering. The larger the value of $k$, the less data is filtered out.

\subsection{Dynamic Strategy}
In our dynamic strategy, we design a loss-aware technique to evaluate the quality of data.

\subsubsection{Dynamic Training Loss}
In contrast to the static strategy, the dynamic strategy integrates loss-aware dynamic evaluations into the data filtering process. This means initiating the process by training a base model and utilizing the loss of individual samples as the primary evaluation metric. Within the domain of deep learning, the loss function serves as a pivotal metric of the model's alignment with the dataset. Hence, a model effectively capturing the underlying knowledge within the data typically yields a lower loss, while ineffective learning manifests in higher loss values.

The advantage of the dynamic strategy lies in its ability to leverage the characteristics of the model, thereby better adapting to the complexity and diversity of the data. We employ the common cross-entropy loss function in deep learning as the approach for calculating loss during model training. 
To better evaluate the performance of the base model, we choose the top three epochs where the model performs best on the validation set to collect the loss values of data samples. We do this for two main reasons: First, the validation set is typically used to assess the model's generalization ability on unseen data. Therefore, we select epochs based on their performance on the validation set to ensure that the collected data samples are representative and possess good generalization ability. Second, models may encounter issues such as overfitting or underfitting during training, resulting in a decrease in their generalization ability. Thus, we choose results from epochs where the model has converged and demonstrated stable performance.

\subsubsection{Dynamic Filter}

\lqy{
In the loss-aware approach, we take into account both the magnitude and fluctuation of the training loss. This dual consideration is important because samples with consistently high loss and significant variance are likely indicative of cases where the model has failed to learn effectively.
For loss magnitude, we simply identify samples whose loss exceeds the upper quartile within an epoch as candidate filtered samples, denoted as $C_1$. For loss fluctuation, we compute the variance of each sample’s loss across multiple epochs. Samples with loss variance above the upper quartile are similarly marked as candidates, denoted as $C_2$.
To fulfill our data cleaning objective, we remove samples that appear in both $C_1$ and $C_2$, as these are most likely to hinder effective model training.
}

\subsection{Model Ensemble}

\lqy{
To integrate the distinct advantages of data cleaned using different strategies, we train two models: $M_s$, based on data filtered by a static strategy, and $M_d$, based on data filtered by a dynamic strategy. We then select the final output by comparing their weighted performance scores, formally defined as:
$$
I = 
\begin{cases}
M_s(x), & \text{if } \alpha \cdot S(M_s) \geq \beta \cdot S(M_d) \\
M_d(x), & \text{otherwise}
\end{cases}
$$
, where $I$ denotes the selected model output for a given input $x$; $\alpha$ and $\beta$ are weights reflecting each model’s performance on a held-out dataset; and $S(\cdot)$ is a scoring function consistent with that used in the static strategy. This approach ensures that the model producing more reliable results under the evaluation metric is chosen at inference time.
}

\section{Experimental setup}
In this section, we first introduce the research questions~(RQs). Then we describe the two datasets for the comment updating task and the metrics for evaluating the effectiveness of the updated comments. Finally, we present baselines and experiment settings.

\subsection{Research Questions}
To evaluate the effectiveness and the efficiency of CupCleaner, we aim to answer the following questions:

\paragraph{RQ1: Whether CupCleaner can filter out true noise data?}
To evaluate CupCleaner's ability to remove noise data, we first record the changes before and after the cleaning process. Then we provide examples that CupCleaner identifies as noisy. Finally, we conduct human evaluation on the noisy and high-quality data identified by CupCleaner to explore whether the filtered data are truly noise.

\paragraph{RQ2: How effective does CupCleaner clean the training data?}

\lqy{
To evaluate the effectiveness of our data cleaning approach, CupCleaner, we conduct experiments on two widely adopted general-purpose code models: PLBART and CodeT5. We apply CupCleaner to clean their training and validation sets, while keeping the test set unchanged across all approaches to ensure fair comparison. This setup ensures that only our approach benefits from data cleaning, while baseline models are trained and validated on noisy data.
We compare the performance of models trained on CupCleaner-cleaned data against multiple baselines. These include both deep learning-based approaches~(e.g., CoditT5) and non-deep learning filtering heuristics that rely on shallow matching of comment and code edits. Through these comparisons, we aim to assess whether CupCleaner can more effectively identify and remove noisy samples, thereby enhancing model performance on the comment updating task.
}

\paragraph{RQ3: How effective is the model ensemble?}
Model ensemble refers to the technique of combining predictions from multiple models to enhance overall performance. To validate the effectiveness of the model ensemble, we first separately evaluate the static and dynamic cleaning strategies. Then, we evaluate the effectiveness of the model ensemble approach by comparing the performance of individual models and the ensembled model on the test set, confirming the impact of individual strategies on the final model performance.


\paragraph{RQ4: How efficient is CupCleaner?}
Calculating the quality score for each data sample involves using pre-trained models and training base models.
This also brings a potential risk that the time required to compute the scores for all data samples may be unacceptable.
To answer this question, we calculate the time consumption for cleaning the entire dataset, including the time required by the static strategy and the dynamic strategy.
We chose the time it takes to train a model for one epoch as a baseline for comparing the time required to clean the data.

\begin{table}[t]
\centering
\renewcommand{\arraystretch}{1.1}
\caption{Table of dataset statistics.}
\scalebox{0.8}{
\begin{tabular}{lll}
\toprule
\textbf{Name}             & \textit{Panthaplackel et al., 2020}  & \textit{Panthaplackel et al., 2021} \\ \midrule
\textbf{Year}             & 2020                                      & 2021                         \\ \midrule
\textbf{Published}        & ACL\footnotemark[1]          & AAAI\footnotemark[2]          \\ \midrule
\textbf{Language}         & Java                                    & Java                         \\ \midrule
\textbf{\#Pairs}          & 7,239                                & 18,522                        \\ \midrule
\textbf{Train/Valid/Test} & 5,791/712/736                     & 16,494/1,878/150              \\ \bottomrule 
\end{tabular}
}
\label{tab_data}
\end{table}
\footnotetext[1]{The Annual Meeting of the Association for Computational Linguistics.}
\footnotetext[2]{The AAAI Conference on Artificial Intelligence.}

\subsection{Datasets}
\lqy{
To evaluate the effectiveness of our data cleaning approach, we conduct experiments on two datasets: \textit{Panthaplackel et al., 2020}\cite{acl20} and \textit{Panthaplackel et al., 2021}\cite{aaai21}.
Table~\ref{tab_data} summarizes the key statistics of these datasets. Both datasets consist of paired old and new comment-code examples that potentially exhibit update relationships. The input to the model includes the old comment, old code, and the modified new code, while the expected output is the updated comment.
}

\lqy{
Specifically, the \textit{Panthaplackel et al., 2020} dataset contains approximately 7.2k samples, collected from the commit histories of open-source Java projects on GitHub, focusing exclusively on changes to \texttt{@return} comments.
In contrast, the \textit{Panthaplackel et al., 2021} dataset is originally constructed for comment consistency detection, and includes a broader range of comment types such as \texttt{@param}, \texttt{@return}, and summary comments. Although not initially designed for comment updating, recent work~\cite{coditt5} has repurposed a subset of this dataset by extracting instances where comment modifications occurred, discarding examples with unchanged comments. The resulting dataset comprises approximately 20k usable samples for comment update research.
}

\lqy{
Both datasets employ simple filtering heuristics to select examples that likely reflect meaningful comment-code updates. For example, the \textit{Panthaplackel et al., 2020} dataset excludes purely stylistic changes (e.g., spelling corrections or reformatting), while the \textit{Panthaplackel et al., 2021} dataset filters out samples with minimal comment edits or code that cannot be parsed into an AST.
However, due to the lack of deeper analysis into the nature of comment updates, both datasets still contain a considerable amount of noise, which motivates the need for more rigorous data cleaning strategies such as the one we propose.
}

\subsection{Metrics}
To evaluate the performance of the models, we select five representative evaluation metrics in the field of text generation and updating, including XMatch, BLEU, METEOR, SARI, and GLEU~\cite{codet5, codebert, coditt5, codereviewer,plbart}.

\subsubsection{XMatch}
The first metric is exact match, i.e., the percentage of generated comments that are identical to the ground truth at the string level.

\subsubsection{BLEU}
The second metric we use is the BLEU score. BLEU\cite{bleu} is originally used to evaluate the performance of machine translation. 
BLEU~\cite{bleu} is commonly used to evaluate the similarity of generated code from a lexical perspective~\cite{lyra, commclean}. Specifically, we use BLEU-4 as the BLEU score.



\subsubsection{METEOR}
Our third metric is METEOR~\cite{meteor}. METEOR utilizes word alignment to align the model-generated results with the reference texts and evaluate the generated results. Compared to BLEU, METEOR does not rely on n-gram matching, but instead places more emphasis on the syntax, word order, and semantics of the generated results.

\subsubsection{SARI}
Our fourth metric is SARI~(System for Automatic Evaluation of Text Simplification). 
SARI~\cite{sari} is an evaluation metric based on edit distance, originally designed to assess the overall quality of text simplification. SARI compares the generated results with the reference in terms of edit distance, as well as the matching of words, phrases, and sentences at various levels, to provide an overall assessment of the quality of the generated results.

\subsubsection{GLEU}
Our last metric is GLEU~\cite{gleu} (Generalized Language Evaluation Understanding). GLEU is a variant of the BLEU metric originally designed for evaluating grammatical correction systems. By rewarding correct edits and penalizing ungrammatical ones, GLEU is closer to human-level judgment than BLEU and is more suitable for tasks involving edits.
GLEU captures fluency and grammatical constraints through the use of n-grams, making it more suitable for evaluating tasks involving edits.

\subsection{Baselines}

\lqy{
To comprehensively evaluate the effectiveness of our proposed method, CupCleaner, we compare it against two types of baselines: (1) widely used pre-trained code models, and (2) existing data cleaning approaches tailored for code-comment tasks. The former comparison aims to assess whether applying CupCleaner can enhance the performance of general-purpose models (e.g., PLBART, CodeT5) beyond their original results on noisy data. The latter comparison focuses on evaluating whether CupCleaner outperforms prior cleaning strategies in identifying high-quality training samples and improving downstream performance. This dual-perspective evaluation allows us to holistically measure the practical impact and general utility of CupCleaner.
}

\subsubsection{Baseline Models}

\lqy{To ensure generalizability, we evaluate CupCleaner using widely adopted and model-agnostic pre-trained code models rather than task-specific architectures. We select three recent and representative models: PLBART\cite{plbart}, CodeT5\cite{codet5}, and CoditT5~\cite{coditt5}.}

\textit{PLBART: }
\lqy{PLBART is a sequence-to-sequence model based on the BART~\cite{bart} architecture and pre-trained on large-scale programming language data, including Java and Python from GitHub and natural language descriptions from StackOverflow. It uses denoising pre-training objectives such as token masking, deletion, and infilling. PLBART has demonstrated strong performance on multiple code understanding and generation tasks, outperforming prior models such as CodeGPT~\cite{codexglue} and CodeBERT~\cite{codebert}.}

\textit{CodeT5: }
\lqy{CodeT5 extends the T5~\cite{t5} encoder-decoder architecture for programming language tasks. It introduces code-aware pre-training objectives, including masked span prediction, identifier tagging, masked identifier prediction, and bimodal dual generation. These tasks help the model better capture the structural and semantic characteristics of code, leading to state-of-the-art results across various code-related tasks~\cite{divot5,twinxsql,codereviewer}.}

\textit{CoditT5: }
\lqy{CoditT5~\cite{coditt5} is built on top of CodeT5 and is tailored specifically for code editing tasks. It employs a denoising pre-training framework that perturbs code snippets based on downstream edit distributions. These perturbations include insertion, deletion, or replacement of tokens. The model learns to recover the original input and predict the editing process. CoditT5 has shown superior performance on annotation update tasks compared to more general models.}

\subsubsection{Baseline Cleaning Approaches}

\lqy{To further validate the effectiveness of CupCleaner, we compare it with two recent code-specific data cleaning strategies that aim to improve training data quality in code-comment tasks.}

\textit{CommClean: }
\lqy{CommClean applies a set of heuristic rules to filter out low-quality code-comment pairs, particularly those with clearly non-informative or automatically generated comments. It is designed to improve the training of comment generation models. When applicable, samples with updated comments are split into independent before/after pairs to enable filtering.}

\textit{OCDData: }
\lqy{OCDData focuses on identifying meaningful comment updates. It first uses heuristic rules to discard samples where the change between two versions of a comment is minimal, assuming that substantial changes indicate more reliable update examples. These filtered samples are then used to train a discriminator to detect outdated comments, with unchanged or unrelated examples treated as negative cases.}

\subsection{Implementation Settings}
The main experimental environment is an Ubuntu OS with two Intel Xeon CPUs, 188GB RAM, and a NVIDIA TITAN RTX GPU with 24GB of memory. 
To save training time and resources, we set the maximum number of epochs for training to 20 and implement an early stopping strategy.
Specifically, we implement a common strategy during the training process, where we terminate the training process if the experimental performance on the validation set does not improve within three consecutive epochs.
For each dataset, we use exactly the same model settings across all variants of the dataset.
In addition, to make the experimental process more generalizable, we simply concatenate the inputs into a long sequence without performing any special processing.
In the static cleaning strategy, to filter out lower-quality data as much as possible, we set the coefficient k to 0.5. In the model ensemble, we utilize the commonly used metric BLEU as the metric for assigning performance weights to each model.

\section{Results}
In this section, we present the experimental results and answer each research question.
\subsection{Noise data identified by CupCleaner~(RQ1)}
To answer this question, we explore the data filtered out by CupCleaner. First, we present the statistical information of data cleaning, then we analyze some samples of low-quality data that CupCleaner identifies, and finally, we conduct a human evaluation of the identified clean data and noisy data.

\begin{table}[t]
\centering
\renewcommand{\arraystretch}{1.5}
\caption{Table of data statistics after cleaning.}
\scalebox{0.6}{
\begin{tabular}{ccccccc}
\hline
\textbf{Dataset}     & \multicolumn{3}{c}{\textit{\textbf{\begin{tabular}[c]{@{}c@{}}Panthaplackel\\ et al., 2020\end{tabular}}}}                                               & \multicolumn{3}{c}{\textbf{\begin{tabular}[c]{@{}c@{}}Panthaplackel\\ et al., 2021\end{tabular}}}                                                 \\ \cline{2-7} 
\multicolumn{1}{l}{} & \multicolumn{1}{l}{\textbf{\# Samples}}                    & \multicolumn{1}{l}{\textbf{\# Old/New Comment Tokens}} & \multicolumn{1}{l}{\textbf{Edit Distance}} & \multicolumn{1}{l}{\textbf{\# Samples}}                     & \multicolumn{1}{l}{\textbf{\# Old/New Comment Tokens}} & \multicolumn{1}{l}{\textbf{Edit Distance}} \\ \hline
Original             & 5,791                                                      & 10.84 / 11.79                                  & 5.99                                       & 16,494                                                      & 13.94 / 14.70                          & 6.27                                       \\ \hline
Static               & \begin{tabular}[c]{@{}c@{}}5,227\\ (-9.74\%)\end{tabular}  & 11.24 / 12.23                                  & 5.65                                       & \begin{tabular}[c]{@{}c@{}}14,922\\ (-9.53\%)\end{tabular}  & 14.31/ 15.13                           & 5.94                                       \\ \hline
Dynamic-PLBART       & \begin{tabular}[c]{@{}c@{}}4,624\\ (-20.15\%)\end{tabular} & 10.47 / 10.17                                  & 4.35                                       & \begin{tabular}[c]{@{}c@{}}12,410\\ (-24.76\%)\end{tabular} & 13.47 / 13.34                          & 4.50                                       \\ \hline
Dynamic-CodeT5       & \begin{tabular}[c]{@{}c@{}}4,764\\ (-17.73\%)\end{tabular} & 11.22 / 11.77                                  & 5.01                                       & \begin{tabular}[c]{@{}c@{}}13,427\\ (-18.59\%)\end{tabular} & 14.20 / 14.27                          & 4.72                                       \\ \hline
\end{tabular}
}
\label{tab_cleaned}
\end{table}

\begin{figure}[t]
\centering
\centerline{\includegraphics[scale=0.75]{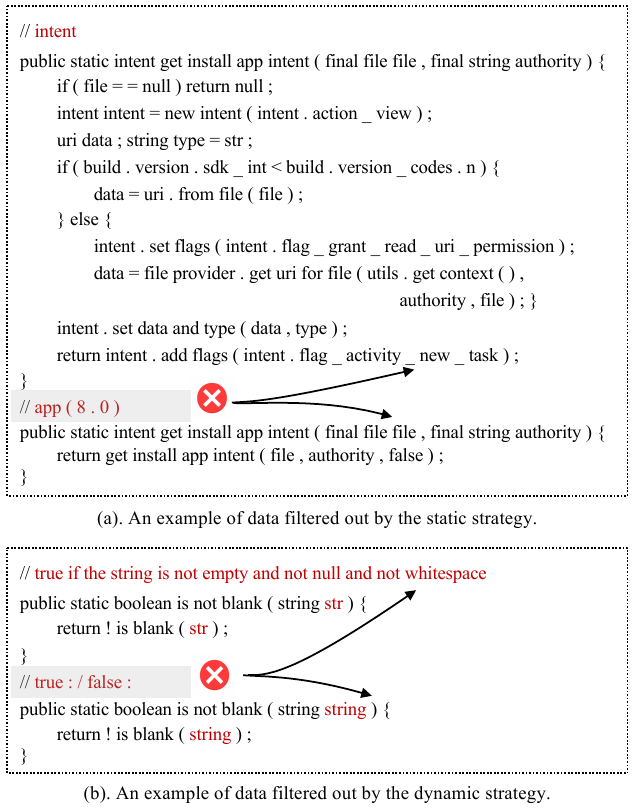}}
\caption{Noise samples detected by the static strategy and the dynamic strategy, respectively.}
\label{fig_noise}
\end{figure}


\lqy{
Table~\ref{tab_cleaned} compares the training data before and after different cleaning strategies across two datasets. The static strategy removes about 9–10\% of samples, but key metrics such as comment length and edit distance remain largely unchanged. This indicates that static rules filter obvious noise while preserving the original data distribution.
In contrast, dynamic strategies show greater variation across models. PLBART filters over 20–24\% of samples, while CodeT5 removes around 18\%, reflecting differences in model capacity and focus. Since dynamic filtering relies on model confidence, the retained samples tend to have clearer and more meaningful edits, leading to a larger drop in average edit distance compared to static filtering.
Therefore, static methods ensure distributional consistency, while dynamic methods favor higher-quality edits, resulting in more pronounced shifts in the dataset.
}

To demonstrate the discriminating ability of CupCleaner on low-quality data, we provide a data sample analysis. Figure~\ref{fig_noise} illustrates the low-quality data samples detected by the static strategy and the dynamic strategy applied on CodeT5, respectively. Subfigure~(a) shows an example where the old code contains many lines, but its comment is inadequate for describing the content of the code. Most of the code lines are removed in the edit from the old code to obtain the new code, but the comments in the new code are weakly related to the comments in the old code and are unsuitable for describing the new code adequately. Therefore, the static strategy filters out this data as low-quality data. 
Subfigure~(b) showcases an example where the dynamic strategy identifies cases where the code content remains largely unchanged, but the comment undergoes significant changes. Moreover, the new comment is deemed ineffective in describing the new code. Hence, such data may mislead the model regarding the modification of old comments, resulting in considerable loss during training.

\lqy{
To further assess whether CupCleaner inadvertently filters out valuable data or introduces distributional bias, we conducted a human evaluation focused on the authenticity of samples removed by the static cleaning strategy.
Three Ph.D. students in computer science volunteered to participate in the evaluation.
This analysis helps verify whether the filtered data indeed lacks semantic alignment between code and comment changes.
}
\lqy{
We first randomly select 50 samples from the high-quality data and 50 samples from the noisy data identified by CupCleaner in the dataset of \textit{Panthaplackel et al., 2020}. 
Due to the susceptibility of the dynamic strategy to different base models, we only investigated whether the data filtered out by the static strategy truly contains noisy data.
Then, we make sure that the evaluators understand the context and the intention of the comment updating task and ask them the question ``Would you accept this data sample into the comment updating dataset?" 
This question means that the annotators need to judge whether the comment update process of each data is reasonable and understandable to humans.
The label for each data sample ranges from 1 to 5, representing reject, weak reject, weak accept, accept, and strong accept, indicating the willingness to accept the data.
}
To ensure objectivity, we assign corresponding characteristics to each score for reference:
\begin{itemize}
    \item Score 1: Comments are mostly unreasonable or non-literal.
    \item Score 2: Comments are generally reasonable, but the changes in comments are unrelated to changes in the code.
    \item Score 3: Comments are reasonable, and the changes in comments are loosely related to changes in the code.
    \item Score 4: Comments are reasonable, and the changes in comments are mostly related to changes in the code.
    \item Score 5: Comments are reasonable, and the changes in comments are strongly related to changes in the code.
\end{itemize}
We gather all volunteer evaluators to collectively review and discuss the scores of each sample. The final score is determined through a voting process involving the participation of three volunteer evaluators.
Finally, the average score of data that CupCleaner considers to be retained is 4.3, while the average score of data filtered out by CupCleaner is 2.5.
This indicates that the majority of the data filtered out by CupCleaner belongs to noisy data, while the data retained is predominantly in line with the intent of the comment updating task. 


 \begin{tcolorbox}
\textbf{Answer to RQ1}: CupCleaner can filter out a significant amount of data (i.e., CodeT5's dynamic strategy filters out approximately 18\% of the data).
The results of the human evaluation show that the average score of the data filtered out by CupCleaner is only 2.5/5, while the average score of the high-quality data identified by CupCleaner as high-quality is 4.3/5. This indicates that CupCleaner can distinguish noisy data from high-quality data in comment updating datasets.
 \end{tcolorbox}

\begin{table}[t]
\centering
\renewcommand{\arraystretch}{1.5}
\caption{Table of experimental results.}

\scalebox{0.72}{
\begin{tabular}{cllllll}
\hline
\textbf{Dataset}                                                                                        & \textbf{Model}    & \textbf{XMatch} & \textbf{BLEU}  & \textbf{Meteor} & \textbf{SARI}  & \textbf{GLEU}  \\ \hline
\multirow{7}{*}{\textit{\textbf{\begin{tabular}[c]{@{}c@{}}Panthaplackel\\ et al., 2020\end{tabular}}}} 
& PLBART            & 12.23           & 39.7           & 35.74           & 39.87          & 34.27          \\
& CodeT5            & 21.60           & 55.42$^\alpha$          & 48.76           & 44.46          & 49.49$^\alpha$          \\
& CoditT5           & 17.39           & 52.65          & 45.68           & 41.64          & 46.39          \\ \cline{2-7} 
& CommClean-CodeT5  & 21.88           & 55.15$^\alpha$          & 48.91           & 45.16          & 49.30          \\
& OCDData-CodeT5    & 21.60           & 55.23$^\alpha$          & 48.81           & 46.02$^\alpha$          & 49.27          \\ \cline{2-7} 
& \textbf{CupCleaner}-PLBART & 16.17           & 48.82          & 43.40           & 42.11          & 42.17          \\
& \textbf{CupCleaner}-CodeT5 & \textbf{23.10}  & \textbf{55.83$^\alpha$} & \textbf{49.54}  & \textbf{46.63$^\alpha$} & \textbf{50.30$^\alpha$} \\ \hline
\multirow{7}{*}{\textit{\textbf{\begin{tabular}[c]{@{}c@{}}Panthaplackel\\ et al., 2021\end{tabular}}}} & PLBART            & 33.33           & 59.68          & 55.70           & 55.23          & 55.08          \\

& CodeT5            & 41.33           & 67.03          & 61.45           & 63.30           & 60.54          \\
& CoditT5           & 43.33           & 64.56          & 60.75           & 61.41          & 59.53          \\ \cline{2-7} 
 & CommClean-CodeT5  & 41.33           & 67.64          & 61.56           & 63.03          & 60.65          \\
 & OCDData-CodeT5    & 40.67           & 67.28          & 60.85           & 63.93          & 59.77          \\ \cline{2-7} 
& \textbf{CupCleaner}-PLBART & 38.00            & 65.06          & 59.24           & 57.37          & 57.50           \\
& \textbf{CupCleaner}-CodeT5 & \textbf{45.33}  & \textbf{69.11} & \textbf{64.37}  & \textbf{65.82} & \textbf{62.52} \\ \hline
\end{tabular}
}
\label{tab_rq1}
\end{table}

\subsection{Effectiveness of training using cleaned data~(RQ2)}
\lqy{Table~\ref{tab_rq1} presents a comprehensive evaluation of our proposed method, CupCleaner, from both the model improvement perspective and the data cleaning effectiveness perspective. This dual-perspective analysis allows us to holistically assess CupCleaner’s practical value.
We use the superscript $\alpha$ to indicate that the difference between baseline and CupCleaner is not statistically significant. Importantly, only our method involves cleaning the training and validation data, while all baseline approaches are trained and validated on the original data. In all experiments, we keep the test set unchanged to ensure fair and consistent evaluation across approaches.
}

\lqy{From the model performance perspective, we aim to evaluate whether applying CupCleaner to general-purpose models~(PLBART, CodeT5) can enhance their performance on noisy real-world data. The results show clear and consistent improvements. On the \textit{Panthaplackel et al., 2020} dataset, CupCleaner-PLBART achieves over 32\% gain in XMatch~(i.e., from 12.23 to 16.17) and over 22\% gain in BLEU (i.e., from 39.7 to 48.82), compared to the original PLBART trained on unfiltered data. Similarly, CupCleaner-CodeT5 improves the already strong CodeT5 model by 6.9\% in XMatch~(i.e., from 21.60 to 23.10) and achieves the highest scores across all metrics. On the \textit{Panthaplackel et al., 2021} dataset, CupCleaner-PLBART improves XMatch by 14.0\%, while CupCleaner-CodeT5 again outperforms all baselines, including CoditT5, with a 4.0 percentage points gain in XMatch, and over 2.0 points gain in BLEU (i.e., from 67.03 to 69.11). For most metrics, the improvements brought by CupCleaner are statistically significant, and even in cases marked with $\alpha$, it still achieves consistent numerical gains over all baselines. These results demonstrate that high-quality training data contributes more significantly to final performance than architectural specialization alone: even a general-purpose model trained on cleaned data can outperform a code-editing-specific model like CoditT5.}

\lqy{From the data cleaning effectiveness perspective, we compare CupCleaner with existing heuristic-based filtering approaches, CommClean and OCDData, applied to CodeT5. As shown in the table, both methods provide minimal or no meaningful gains. On the \textit{Panthaplackel et al., 2020} dataset, CommClean improves XMatch only marginally (i.e., from 21.60 to 21.88), while BLEU drops slightly~(i.e., from 55.42 to 55.15). OCDData performs similarly, with unchanged XMatch and decreased BLEU~(55.23). On the \textit{Panthaplackel et al., 2021} dataset, CommClean and OCDData improve BLEU by less than 1 point, and in some cases, GLEU and SARI even drop. In contrast, CupCleaner not only achieves consistently higher scores on all metrics but also introduces a more principled and scalable cleaning mechanism: a static filter based on semantic rules and a dynamic filter guided by model loss.}
\lqy{
According to our further analysis, CommClean and OCDData each remove only about 1\% of the data, and in some cases, as little as 0.26\% (e.g., by CommClean on the second dataset). By comparison, CupCleaner filters out over 9\% of the data through its static filtering mechanism, and removes over 17\% of noisy samples via its dynamic loss-aware strategy. This demonstrates that existing heuristic rules are too conservative to identify noisy training data effectively, especially in the context of comment updates.
}

\lqy{In summary, these results confirm that CupCleaner is both an effective data quality enhancer and a robust training-time plug-in for improving code-comment models. Its ability to deliver consistent improvements across multiple datasets, models, and metrics highlights its generality, while its clear outperformance of prior cleaning methods underscores its practical utility in real-world noisy-code scenarios.}

\begin{tcolorbox}
\textbf{Answer to RQ2}:  
\lqy{
Models trained on data cleaned with CupCleaner consistently outperform those trained on unfiltered data, surpassing even specialized models like CoditT5 across all metrics. For example, on the \textit{Panthaplackel et al., 2021} dataset, the BLEU score of PLBART improves from 59.68\% to 65.06\%, and CodeT5 improves from 67.03\% to 69.11\%. Compared to existing cleaning methods like CommClean and OCDData, which filter less than 1\% of data and yield marginal gains, CupCleaner removes over 9\% of noisy samples and delivers significant performance improvements. These results confirm CupCleaner’s effectiveness in enhancing data quality and model performance in comment updating tasks.
}
\end{tcolorbox}

\begin{table}[t]
\centering
\renewcommand{\arraystretch}{1.5}
\caption{Table of model ensemble.}
\scalebox{0.8}{
\begin{tabular}{ccllllll}
\hline
\textbf{Dataset}                   & \multicolumn{1}{l}{} & \multicolumn{3}{c}{\textit{\textbf{\begin{tabular}[c]{@{}c@{}}Panthaplackel\\ et al., 2020\end{tabular}}}} & \multicolumn{3}{c}{\textit{\textbf{\begin{tabular}[c]{@{}c@{}}Panthaplackel\\ et al., 2021\end{tabular}}}} \\ \hline
\textbf{Model}                     & \multicolumn{1}{l}{} & \textbf{XMatch}                    & \textbf{BLEU}                     & \textbf{Meteor}                   & \textbf{XMatch}                    & \textbf{BLEU}                     & \textbf{Meteor}                   \\ \hline
\multirow{3}{*}{CupCleaner-PLBART} & Ensemble             & \textbf{16.17}                     & \textbf{48.82}                    & \textbf{43.4}                     & \textbf{38.00}                     & \textbf{65.06}                    & \textbf{59.24}                    \\ \cline{2-8} 
                                   & Static               & 14.67                              & 45.73                             & 41.12                             & 36.00                               & 62.87                             & 57.46                            \\ \cline{2-8} 
                                   & Dynamic              & 14.81                              & 44.45                             & 39.70                              & 38.00                              & 63.33                             & 58.91                             \\ \hline
\multirow{3}{*}{CupCleaner-CodeT5} & Ensemble             & 23.10                               & \textbf{55.83}                    & 49.54                             & \textbf{45.33}                     & \textbf{69.11}                    & \textbf{64.37}                    \\ \cline{2-8} 
                                   & Static               & \textbf{23.51}                     & 55.78                             & \textbf{49.71}                    & 44.00                               & 67.89                             & 63.28                             \\ \cline{2-8} 
                                   & Dynamic              & 22.15                              & 54.46                             & 48.59                             & 44.67                              & 68.98                             & 63.83                             \\ \hline
\end{tabular}
}
\label{tab_rq2}
\end{table}

\subsection{Effectiveness of model ensemble~(RQ3)}
\lqy{To answer RQ3, we conduct an ablation study to evaluate the individual impact of the static and dynamic cleaning strategies, as well as the effect of their ensemble. Table~\ref{tab_rq2} reports the results on both datasets, highlighting the first three representative evaluation metrics~(XMatch, BLEU, and Meteor), with other metrics showing similar trends.}

\lqy{We observe that both static and dynamic strategies lead to performance improvements over models trained on the original unfiltered data. For example, with the CupCleaner-PLBART model on the \textit{Panthaplackel et al., 2020} dataset, static and dynamic strategies improve XMatch from 12.23\% to 14.67\% and 14.81\%, respectively. On the \textit{Panthaplackel et al., 2021} dataset, static cleaning boosts XMatch from 33.33\% to 36.00\%, while dynamic cleaning raises it further to 38.00\%. Similar trends are observed for CupCleaner-CodeT5: on the 2020 dataset, static cleaning yields slightly better XMatch (23.51\%) than dynamic (22.15\%); whereas on the 2021 dataset, dynamic cleaning (44.67\%) slightly outperforms static (44.00\%).}

\lqy{While the relative effectiveness of the two strategies may vary by dataset and model, the ensemble approach generally offers the best overall performance. For example, CupCleaner-CodeT5 on the 2021 dataset achieves the highest XMatch (45.33\%), BLEU (69.11\%), and Meteor (64.37\%) when combining both strategies. CupCleaner-PLBART also benefits from ensembling, with BLEU improving from 63.33\% (dynamic only) to 65.06\%, despite XMatch remaining the same. These results indicate that the ensemble mechanism effectively integrates complementary strengths from both static and dynamic strategies.}

\lqy{Notably, the ensemble selection mechanism adopts the same criteria as used in the static cleaning strategy to rank candidate outputs. This consistency further reinforces the effectiveness and reliability of the static component.}

\begin{tcolorbox}
\textbf{Answer to RQ3:}
\lqy{Both static and dynamic cleaning strategies individually improve model performance across datasets. Their combination through model ensemble further enhances generalization by leveraging the complementary benefits of each strategy. For example, on the \textit{Panthaplackel et al., 2021} dataset, CupCleaner-CodeT5 achieves higher scores across all metrics with ensembling (e.g., BLEU improves from 67.89\% to 69.11\%, and Meteor from 63.28\% to 64.37\%), demonstrating the effectiveness of the combined approach.}
\end{tcolorbox}

\subsection{Time consumption of data cleaning process~(RQ4)}

\begin{table}[t]
\centering
\renewcommand{\arraystretch}{1.5}
\caption{Table of the time consumption.}
\scalebox{0.8}{
\begin{tabular}{lll}
\toprule
                   & \textit{\begin{tabular}[c]{@{}l@{}}Panthaplackel et al,\\ et al., 2020\end{tabular}} & \textit{\begin{tabular}[c]{@{}l@{}}Panthaplackel et al, \\ et al., 2021\end{tabular}} \\ \midrule
Training one epoch & 1 epoch (4.38 minutes)                                                       & 1 epoch (12.15 minutes)                                                       \\ \midrule
Static             & $\approx$ 0.5 epoch                                                     &  $\approx$ 0.5 epoch                                                       \\ \midrule
Dynamic            &  $\approx$ n epochs                                                        & $\approx$ n epochs                                                         \\ \bottomrule
\end{tabular}
}
\label{tab_rq3}
\end{table}

\lqy{
To answer RQ4, we examine the time required to clean the dataset and compare it to training one epoch of the model on the raw dataset. 
An epoch means that the model goes through all the training data once and adjusts its weights based on the loss function during the training process.
We select CodeT5 as the baseline model to calculate the time required to train one epoch, as it has exhibited relatively high performance in previous experiments.
In one epoch of model training, we calculate the time spent on the training set and the time spent on the validation set after one epoch.
During the data cleaning process, we calculate the time spent on static and dynamic strategies.
Table~\ref{tab_rq3} shows the basic statistics of time consumption, where training one epoch on the \textit{Panthaplackel et al., 2020} dataset takes approximately 4.38 minutes, while on the \textit{Panthaplackel et al., 2021} dataset, it takes around 12.15 minutes.
}


\lqy{
For the static cleaning strategy, the total time consumption is approximately equivalent to 0.5 training epochs, around 2.2 minutes on the \textit{Panthaplackel et al., 2020} dataset and 6.7 minutes on the \textit{Panthaplackel et al., 2021} dataset. This is because generating semantic representations for all training samples only involves the encoder, without requiring full forward-backward propagation, effectively halving the computational workload compared to a full epoch. After this encoding step, the filtering process itself is lightweight and incurs negligible additional time.
}

\lqy{
The dynamic cleaning strategy requires training the model to converge in order to collect loss signals for filtering, which involves n epochs of training. However, this cost is typically amortized or unavoidable in practice, as model training is necessary regardless of whether cleaning is applied. Moreover, the cost is negligible when compared to pre-training: for instance, CoditT5 was pre-trained for 4 days~\cite{coditt5}, while fine-tuning on the 2020 dataset took only 35 minutes. The loss analysis step itself incurs minimal overhead~(i.e., about 0.1 minutes).
}

\lqy{
In summary, both the static and dynamic strategies are efficient in practice. Static cleaning is lightweight, and dynamic cleaning reuses already-required training steps. Therefore, the modest additional time required by CupCleaner is acceptable and well justified by the significant improvements it brings to data quality and model performance.
}

\begin{tcolorbox}
\textbf{Answer to RQ4}: 
\lqy{
The time cost of CupCleaner is low and practical. Static cleaning requires less time than a single training epoch, while dynamic cleaning leverages standard training steps and adds negligible overhead. Overall, the modest time investment is well justified by the performance and generalization improvements enabled by higher-quality training data.
}
\end{tcolorbox}


\section{Discussion}

\paragraph{Threshold Selection and the Role of Parameter $k$.}
\lqy{In our static semantic-aware filtering strategy, we adopt a thresholding mechanism inspired by the box-plot statistical method. Specifically, we filter out data samples whose semantic similarity scores fall below $Q_1 - k \cdot IQR$, where $Q_1$ is the lower quartile and $IQR$ is the interquartile range. This design is motivated by the classical use of box plots to identify statistical outliers. Using $Q_1$ instead of the median or mean offers robustness to skewed distributions, which are common in real-world code-comment similarity scores, and allows us to focus on the ``low-quality tail'' of the distribution more effectively.}

\lqy{The parameter $k$ controls the aggressiveness of the filtering strategy: larger values of $k$ result in more conservative thresholds (removing fewer samples), while smaller values make the strategy more aggressive. We conducted further experiments to evaluate the sensitivity of CupCleaner to different $k$ values. The results show that CupCleaner consistently improves model performance across a range of $k$ values (e.g., 0.5, 1.0, 1.5), indicating that the method is robust to the specific choice of threshold.
For example, on the \textit{Panthaplackel et al., 2021} dataset, the original CodeT5 model~(without cleaning) achieves an XMatch score of 41.33\% and a BLEU score of 67.03\%. When applying CupCleaner with $k = 1.0$, the model improves to 42.00\% (XMatch) and 67.49\% (BLEU). When the threshold is set to a more conservative value with $k = 1.5$, the scores further increase to 43.33\% (XMatch) and 68.44\% (BLEU). These results demonstrate that CupCleaner consistently enhances model performance under different filtering intensities, confirming the robustness and practical value of our strategy. In practice, $k=0.5$ offers a balanced trade-off between recall and precision in identifying noisy samples, which is why we adopt it as the default setting.}

\paragraph{Generalizability to Other Programming Languages.}
\lqy{Our current evaluation focuses on Java datasets, which aligns with prior work in comment updating~\cite{acl20,aaai21} and reflects Java’s wide adoption in open-source projects. Nonetheless, CupCleaner is language-agnostic by design: its semantic filtering relies on pretrained models' representations of code-comment pairs, and its dynamic strategy is guided by model loss, both of which apply to other programming languages as long as pretrained encoders are available. While we leave multi-language evaluation for future work, we believe our method can generalize well to other languages, such as Python or JavaScript, especially with the growing availability of multilingual code models.}

\section{Threats to Validity}

\textbf{Threats to internal validity} may arise from three factors. The first threat relates to the implementation of our experiments. 
To mitigate this threat, we load pre-trained model parameters using the HuggingFace~\cite{huggingface} framework and employ its standard library to construct the training and evaluation framework for the model. Additionally, we set a random seed to ensure the reproducibility of experiments. Finally, the implementation of model evaluation adopted in our study remains consistent with the existing literature.
The second threat is that while filtering out noisy data, we might inadvertently exclude normal data as well. This is easily understandable because defining whether a data sample is noise can be challenging. On one hand, programmers have different ideas and styles when coding, and on the other hand, many data samples require an in-depth understanding to determine their appropriateness. 
To mitigate this threat, we employ various evaluation metrics from the code generation domain to assess whether models trained on cleaned data can effectively enhance performance and generalization. Experimental results indicate that our cleaning approach significantly improves \lqy{improves the quality of the training set} and enhances multiple model metrics.
The third threat relates to human evaluation. In our human evaluation, we present annotators with only one question and ask them to provide ratings based on their responses to this question. A comprehensive assessment of a data sample may require analysis from multiple perspectives. However, the annotators we invite are all Ph.D students with relevant development experience. To ensure their understanding of the comment updating task, we also engage in discussions during the annotation process to mitigate the potential impact of this threat.

\textbf{Threat to external validity} may come from the choice of the baseline models. In our experiments, we exclusively choose pre-trained models and do not include non-pre-trained models, which could potentially introduce bias into the final experimental results. We assume that pre-trained models possess a broader base of knowledge, making them more capable of understanding downstream tasks compared to training from scratch. Actually, pre-trained models typically exhibit better performance and are also employed in various generative tasks.

\section{Conclusion}
In this paper, we propose CupCleaner, a data cleaning approach for comment updating datasets. 
Our data cleaning approach combines a semantic-aware static cleaning strategy with a loss-aware dynamic cleaning strategy.
In the experiments to evaluate CupCleaner, we first conduct a human evaluation, and statistical results suggest that the data identified as noise by CupCleaner receive significantly low scores.
Then, we compare models trained on data cleaned with CupCleaner to other code pre-training models. Results show that models trained on cleaned data significantly outperform those trained on original data on the same test set. 
We also compare the experimental performance of different cleaning strategies and model ensembles. Results indicate that while different cleaning strategies exhibit varied performances on different datasets, they all show improvement compared to models trained on raw data, demonstrating the effectiveness of each cleaning strategy. 
Furthermore, the model ensemble effectively leverages the characteristics of different strategies, further enhancing model performance and generalization. Finally, we investigate the time consumption of data cleaning and find that the time spent on cleaning data is acceptable compared to fine-tuning models.

In future work, we plan to explore other datasets for various software engineering tasks and expand our data cleaning approach.

\section{Acknowledgment}
This work is sponsored by the National Key Research and Development Program of China under Grant No. 2023YFB4503803, the National Natural Science Foundation of China under Grant No. 62232003 and No. 62402482.








\printbibliography

@article{acl20,
  title={Learning to update natural language comments based on code changes},
  author={Panthaplackel, Sheena and Nie, Pengyu and Gligoric, Milos and Li, Junyi Jessy and Mooney, Raymond J},
  journal={arXiv preprint arXiv:2004.12169},
  year={2020}
}

@inproceedings{ase20,
  title={Automating just-in-time comment updating},
  author={Liu, Zhongxin and Xia, Xin and Yan, Meng and Li, Shanping},
  booktitle={Proceedings of the 35th IEEE/ACM International Conference on Automated Software Engineering},
  pages={585--597},
  year={2020}
}

@inproceedings{aaai21,
  title={Deep just-in-time inconsistency detection between comments and source code},
  author={Panthaplackel, Sheena and Li, Junyi Jessy and Gligoric, Milos and Mooney, Raymond J},
  booktitle={Proceedings of the AAAI Conference on Artificial Intelligence},
  volume={35},
  number={1},
  pages={427--435},
  year={2021}
}

@inproceedings{bleu,
    author = {Papineni, Kishore and Roukos, Salim and Ward, Todd and Zhu, Wei-Jing},
    title = {BLEU: A Method for Automatic Evaluation of Machine Translation},
    booktitle = {ACL},
    year = {2002},
    url = {https://doi.org/10.3115/1073083.1073135},
    doi = {10.3115/1073083.1073135},
    pages = {311–318},
    numpages = {8},
    location = {Philadelphia, Pennsylvania}
}

@inproceedings{meteor,
  title={METEOR: An automatic metric for MT evaluation with improved correlation with human judgments},
  author={Banerjee, Satanjeev and Lavie, Alon},
  booktitle={Proceedings of the acl workshop on intrinsic and extrinsic evaluation measures for machine translation and/or summarization},
  pages={65--72},
  year={2005}
}

@article{sari,
  title={Optimizing statistical machine translation for text simplification},
  author={Xu, Wei and Napoles, Courtney and Pavlick, Ellie and Chen, Quanze and Callison-Burch, Chris},
  journal={Transactions of the Association for Computational Linguistics},
  volume={4},
  pages={401--415},
  year={2016},
  publisher={MIT Press}
}

@inproceedings{gleu,
  title={Ground truth for grammatical error correction metrics},
  author={Napoles, Courtney and Sakaguchi, Keisuke and Post, Matt and Tetreault, Joel},
  booktitle={Proceedings of the 53rd Annual Meeting of the Association for Computational Linguistics and the 7th International Joint Conference on Natural Language Processing (Volume 2: Short Papers)},
  pages={588--593},
  year={2015}
}

@article{bert,
  title={Bert: Pre-training of deep bidirectional transformers for language understanding},
  author={Devlin, Jacob and Chang, Ming-Wei and Lee, Kenton and Toutanova, Kristina},
  journal={arXiv preprint arXiv:1810.04805},
  year={2018}
}

@article{roberta,
  title={Roberta: A robustly optimized bert pretraining approach},
  author={Liu, Yinhan and Ott, Myle and Goyal, Naman and Du, Jingfei and Joshi, Mandar and Chen, Danqi and Levy, Omer and Lewis, Mike and Zettlemoyer, Luke and Stoyanov, Veselin},
  journal={arXiv preprint arXiv:1907.11692},
  year={2019}
}

@article{codebert,
  title={Codebert: A pre-trained model for programming and natural languages},
  author={Feng, Zhangyin and Guo, Daya and Tang, Duyu and Duan, Nan and Feng, Xiaocheng and Gong, Ming and Shou, Linjun and Qin, Bing and Liu, Ting and Jiang, Daxin and others},
  journal={arXiv preprint arXiv:2002.08155},
  year={2020}
}

@article{gcb,
  title={Graphcodebert: Pre-training code representations with data flow},
  author={Guo, Daya and Ren, Shuo and Lu, Shuai and Feng, Zhangyin and Tang, Duyu and Liu, Shujie and Zhou, Long and Duan, Nan and Svyatkovskiy, Alexey and Fu, Shengyu and others},
  journal={arXiv preprint arXiv:2009.08366},
  year={2020}
}

@article{plbart,
  title={Unified pre-training for program understanding and generation},
  author={Ahmad, Wasi Uddin and Chakraborty, Saikat and Ray, Baishakhi and Chang, Kai-Wei},
  journal={arXiv preprint arXiv:2103.06333},
  year={2021}
}

@article{unixcoder,
  title={Unixcoder: Unified cross-modal pre-training for code representation},
  author={Guo, Daya and Lu, Shuai and Duan, Nan and Wang, Yanlin and Zhou, Ming and Yin, Jian},
  journal={arXiv preprint arXiv:2203.03850},
  year={2022}
}

@article{codet5,
  title={Codet5: Identifier-aware unified pre-trained encoder-decoder models for code understanding and generation},
  author={Wang, Yue and Wang, Weishi and Joty, Shafiq and Hoi, Steven CH},
  journal={arXiv preprint arXiv:2109.00859},
  year={2021}
}

@article{bart,
  title={Bart: Denoising sequence-to-sequence pre-training for natural language generation, translation, and comprehension},
  author={Lewis, Mike and Liu, Yinhan and Goyal, Naman and Ghazvininejad, Marjan and Mohamed, Abdelrahman and Levy, Omer and Stoyanov, Ves and Zettlemoyer, Luke},
  journal={arXiv preprint arXiv:1910.13461},
  year={2019}
}

@article{lyra,
  title={Lyra: A Benchmark for Turducken-Style Code Generation},
  author={Liang, Qingyuan and Sun, Zeyu and Zhu, Qihao and Zhang, Wenjie and Yu, Lian and Xiong, Yingfei and Zhang, Lu},
  journal={arXiv preprint arXiv:2108.12144},
  year={2021}
}

@inproceedings{commclean,
  title={Are we building on the rock? on the importance of data preprocessing for code summarization},
  author={Shi, Lin and Mu, Fangwen and Chen, Xiao and Wang, Song and Wang, Junjie and Yang, Ye and Li, Ge and Xia, Xin and Wang, Qing},
  booktitle={Proceedings of the 30th ACM Joint European Software Engineering Conference and Symposium on the Foundations of Software Engineering},
  pages={107--119},
  year={2022}
}

@article{t5,
  title={Exploring the limits of transfer learning with a unified text-to-text transformer},
  author={Raffel, Colin and Shazeer, Noam and Roberts, Adam and Lee, Katherine and Narang, Sharan and Matena, Michael and Zhou, Yanqi and Li, Wei and Liu, Peter J},
  journal={The Journal of Machine Learning Research},
  volume={21},
  number={1},
  pages={5485--5551},
  year={2020},
  publisher={JMLRORG}
}

@article{codexglue,
  title={Codexglue: A machine learning benchmark dataset for code understanding and generation},
  author={Lu, Shuai and Guo, Daya and Ren, Shuo and Huang, Junjie and Svyatkovskiy, Alexey and Blanco, Ambrosio and Clement, Colin and Drain, Dawn and Jiang, Daxin and Tang, Duyu and others},
  journal={arXiv preprint arXiv:2102.04664},
  year={2021}
}

@inproceedings{recoder,
  title={A syntax-guided edit decoder for neural program repair},
  author={Zhu, Qihao and Sun, Zeyu and Xiao, Yuan-an and Zhang, Wenjie and Yuan, Kang and Xiong, Yingfei and Zhang, Lu},
  booktitle={Proceedings of the 29th ACM Joint Meeting on European Software Engineering Conference and Symposium on the Foundations of Software Engineering},
  pages={341--353},
  year={2021}
}

@article{aidata,
  title={Data-centric artificial intelligence: A survey},
  author={Zha, Daochen and Bhat, Zaid Pervaiz and Lai, Kwei-Herng and Yang, Fan and Jiang, Zhimeng and Zhong, Shaochen and Hu, Xia},
  journal={arXiv preprint arXiv:2303.10158},
  year={2023}
}

@inproceedings{coditt5,
  title={CoditT5: Pretraining for Source Code and Natural Language Editing},
  author={Zhang, Jiyang and Panthaplackel, Sheena and Nie, Pengyu and Li, Junyi Jessy and Gligoric, Milos},
  booktitle={37th IEEE/ACM International Conference on Automated Software Engineering},
  pages={1--12},
  year={2022}
}

@inproceedings{todocomm,
  title={Automating the removal of obsolete TODO comments},
  author={Gao, Zhipeng and Xia, Xin and Lo, David and Grundy, John and Zimmermann, Thomas},
  booktitle={Proceedings of the 29th ACM Joint Meeting on European Software Engineering Conference and Symposium on the Foundations of Software Engineering},
  pages={218--229},
  year={2021}
}

@article{codereviewer,
  title={CodeReviewer: Pre-Training for Automating Code Review Activities},
  author={Li, Zhiyu and Lu, Shuai and Guo, Daya and Duan, Nan and Jannu, Shailesh and Jenks, Grant and Majumder, Deep and Green, Jared and Svyatkovskiy, Alexey and Fu, Shengyu and others},
  journal={arXiv preprint arXiv:2203.09095},
  year={2022}
}

@inproceedings{repair1,
  title={Shaping program repair space with existing patches and similar code},
  author={Jiang, Jiajun and Xiong, Yingfei and Zhang, Hongyu and Gao, Qing and Chen, Xiangqun},
  booktitle={Proceedings of the 27th ACM SIGSOFT international symposium on software testing and analysis},
  pages={298--309},
  year={2018}
}

@inproceedings{repair2,
  title={Automatic patch generation by learning correct code},
  author={Long, Fan and Rinard, Martin},
  booktitle={Proceedings of the 43rd Annual ACM SIGPLAN-SIGACT Symposium on Principles of Programming Languages},
  pages={298--312},
  year={2016}
}

@inproceedings{repair3,
  title={Context-aware patch generation for better automated program repair},
  author={Wen, Ming and Chen, Junjie and Wu, Rongxin and Hao, Dan and Cheung, Shing-Chi},
  booktitle={Proceedings of the 40th international conference on software engineering},
  pages={1--11},
  year={2018}
}

@inproceedings{repair4,
  title={Learning to synthesize},
  author={Xiong, Yingfei and Wang, Bo and Fu, Guirong and Zang, Linfei},
  booktitle={Proceedings of the 4th International Workshop on Genetic Improvement Workshop},
  pages={37--44},
  year={2018}
}

@inproceedings{repair5,
  title={Precise condition synthesis for program repair},
  author={Xiong, Yingfei and Wang, Jie and Yan, Runfa and Zhang, Jiachen and Han, Shi and Huang, Gang and Zhang, Lu},
  booktitle={2017 IEEE/ACM 39th International Conference on Software Engineering (ICSE)},
  pages={416--426},
  year={2017},
  organization={IEEE}
}

@inproceedings{codereview1,
  title={Expectations, outcomes, and challenges of modern code review},
  author={Bacchelli, Alberto and Bird, Christian},
  booktitle={2013 35th International Conference on Software Engineering (ICSE)},
  pages={712--721},
  year={2013},
  organization={IEEE}
}

@inproceedings{codereivew2,
  title={Modern code review: a case study at google},
  author={Sadowski, Caitlin and S{\"o}derberg, Emma and Church, Luke and Sipko, Michal and Bacchelli, Alberto},
  booktitle={Proceedings of the 40th international conference on software engineering: Software engineering in practice},
  pages={181--190},
  year={2018}
}

@inproceedings{codesearch,
  title={On the importance of building high-quality training datasets for neural code search},
  author={Sun, Zhensu and Li, Li and Liu, Yan and Du, Xiaoning and Li, Li},
  booktitle={Proceedings of the 44th International Conference on Software Engineering},
  pages={1609--1620},
  year={2022}
}

@article{codecomment,
  title={Does your code need comment?},
  author={Huang, Yuan and Jia, Nan and Shu, Junhuai and Hu, Xinyu and Chen, Xiangping and Zhou, Qiang},
  journal={Software: Practice and Experience},
  volume={50},
  number={3},
  pages={227--245},
  year={2020},
  publisher={Wiley Online Library}
}

@inproceedings{codesummary,
  title={On the evaluation of neural code summarization},
  author={Shi, Ensheng and Wang, Yanlin and Du, Lun and Chen, Junjie and Han, Shi and Zhang, Hongyu and Zhang, Dongmei and Sun, Hongbin},
  booktitle={Proceedings of the 44th International Conference on Software Engineering},
  pages={1597--1608},
  year={2022}
}

@article{githubnoise1,
  title={Curating github for engineered software projects},
  author={Munaiah, Nuthan and Kroh, Steven and Cabrey, Craig and Nagappan, Meiyappan},
  journal={Empirical Software Engineering},
  volume={22},
  pages={3219--3253},
  year={2017},
  publisher={Springer}
}

@inproceedings{githubnoise2,
  title={A data set for social diversity studies of GitHub teams},
  author={Vasilescu, Bogdan and Serebrenik, Alexander and Filkov, Vladimir},
  booktitle={2015 IEEE/ACM 12th working conference on mining software repositories},
  pages={514--517},
  year={2015},
  organization={IEEE}
}

@inproceedings {githubnoise3,
    author = {Md Omar Faruk Rokon and Risul Islam and Ahmad Darki and Evangelos E. Papalexakis and Michalis Faloutsos},
    title = {{SourceFinder}: Finding Malware {Source-Code} from Publicly Available Repositories in {GitHub}},
    booktitle = {23rd International Symposium on Research in Attacks, Intrusions and Defenses (RAID 2020)},
    year = {2020},
    isbn = {978-1-939133-18-2},
    address = {San Sebastian},
    pages = {149--163},
    url = {https://www.usenix.org/conference/raid2020/presentation/omar},
    publisher = {USENIX Association},
    month = oct,
}

@inproceedings{conala,
  title={Learning to mine aligned code and natural language pairs from stack overflow},
  author={Yin, Pengcheng and Deng, Bowen and Chen, Edgar and Vasilescu, Bogdan and Neubig, Graham},
  booktitle={Proceedings of the 15th international conference on mining software repositories},
  pages={476--486},
  year={2018}
}

@article{huggingface,
   author = {Huggingface},
   title = {Huggingface Transformers},
   year={2023},
   url = { https://huggingface.co}
}

@article{lientz1978softwaremainten,
  title={Characteristics of application software maintenance},
  author={Lientz, Bennet P and Swanson, E. Burton and Tompkins, Gail E},
  journal={Communications of the ACM},
  volume={21},
  number={6},
  pages={466--471},
  year={1978},
  publisher={ACM New York, NY, USA}
}

@inproceedings{bennett2000softwareevolution,
  title={Software maintenance and evolution: a roadmap},
  author={Bennett, Keith H and Rajlich, V{\'a}clav T},
  booktitle={Proceedings of the Conference on the Future of Software Engineering},
  pages={73--87},
  year={2000}
}

@inproceedings{tufano2021codereview,
  title={Towards automating code review activities},
  author={Tufano, Rosalia and Pascarella, Luca and Tufano, Michele and Poshyvanyk, Denys and Bavota, Gabriele},
  booktitle={2021 IEEE/ACM 43rd International Conference on Software Engineering (ICSE)},
  pages={163--174},
  year={2021},
  organization={IEEE}
}

@inproceedings{just2014defects4j,
  title={Defects4J: A database of existing faults to enable controlled testing studies for Java programs},
  author={Just, Ren{\'e} and Jalali, Darioush and Ernst, Michael D},
  booktitle={Proceedings of the 2014 international symposium on software testing and analysis},
  pages={437--440},
  year={2014}
}

@inproceedings{ocddata_quality,
  title={Data Quality Matters: A Case Study of Obsolete Comment Detection},
  author={Xu, Shengbin and Yao, Yuan and Xu, Feng and Gu, Tianxiao and Xu, Jingwei and Ma, Xiaoxing},
  booktitle={2023 IEEE/ACM 45th International Conference on Software Engineering (ICSE)},
  pages={781--793},
  year={2023},
  organization={IEEE}
}

@article{cupcleaner,
   author = {CupCleaner},
   title = {https://github.com/LIANGQINGYUAN/CupCleaner\_Hybrid},
   year={2025}
}

@article{scis_deeplearning,
author={Chen, Xiangping and Hu, Xing and Huang, Yuan and Jiang, He and Ji, Weixing and Jiang, Yanjie and Jiang, Yanyan and Liu, Bo and Liu, Hui and Li, Xiaochen and others},
  title = {Deep learning-based software engineering: progress, challenges, and opportunities},
  journal = "SCIENCE CHINA Information Sciences",
  year = "2025",
  volume = "68",
  number = "1",
  pages = "111102-",
  url = "http://www.sciengine.com/publisher/Science China Press/journal/SCIENCE CHINA Information Sciences/68/1/10.1007/s11432-023-4127-5",
  doi = "https://doi.org/10.1007/s11432-023-4127-5"
}

@article{twinxsql,
author = {Liang, Qingyuan and Sun, Zeyu and Zhao, Yifan and Gong, Zhihao and Wang, Guoqing and Chen, Yizhou and Zhang, Lu and Liang, Guangtai and Wang, Qianxiang},
title = {Bipartite-Grammar Aware Pretraining for XML-SQL Code Updating},
year = {2025},
publisher = {Association for Computing Machinery},
address = {New York, NY, USA},
issn = {1049-331X},
url = {https://doi.org/10.1145/3731752},
doi = {10.1145/3731752},
note = {Just Accepted},
journal = {ACM Trans. Softw. Eng. Methodol.},
month = apr
}

@misc{grammarcoder,
      title={Grammar-Based Code Representation: Is It a Worthy Pursuit for LLMs?}, 
      author={Qingyuan Liang and Zhao Zhang and Zeyu Sun and Zheng Lin and Qi Luo and Yueyi Xiao and Yizhou Chen and Yuqun Zhang and Haotian Zhang and Lu Zhang and Bin Chen and Yingfei Xiong},
      year={2025},
      eprint={2503.05507},
      archivePrefix={arXiv},
      primaryClass={cs.PL},
      url={https://arxiv.org/abs/2503.05507}, 
}

@article{divot5,
  title={Directional Diffusion-Style Code Editing Pre-training},
  author={Liang, Qingyuan and Sun, Zeyu and Zhu, Qihao and Hu, Junhao and Zhao, Yifan and Chen, Yizhou and Zhu, Mingxuan and Wang, Guoqing and Zhang, Lu},
  journal={arXiv preprint arXiv:2501.12079},
  year={2025}
}

@article{wang2024advanced,
  title={Do advanced language models eliminate the need for prompt engineering in software engineering?},
  author={Wang, Guoqing and Sun, Zeyu and Gong, Zhihao and Ye, Sixiang and Chen, Yizhou and Zhao, Yifan and Liang, Qingyuan and Hao, Dan},
  journal={arXiv preprint arXiv:2411.02093},
  year={2024}
}

@article{liang2024condor,
  title={Condor: A Code Discriminator Integrating General Semantics with Code Details},
  author={Liang, Qingyuan and Zhang, Zhao and Liu, Chen and Sun, Zeyu and Zhang, Wenjie and Chen, Yizhou and Zhao, Zixiao and Luo, Qi and Wang, Wentao and Jiang, Yanjie and others},
  journal={arXiv preprint arXiv:2412.17429},
  year={2024}
}

@article{guo2024deepseekcoder,
  title={DeepSeek-Coder: When the Large Language Model Meets Programming--The Rise of Code Intelligence},
  author={Guo, Daya and Zhu, Qihao and Yang, Dejian and Xie, Zhenda and Dong, Kai and Zhang, Wentao and Chen, Guanting and Bi, Xiao and Wu, Yu and Li, YK and others},
  journal={arXiv preprint arXiv:2401.14196},
  year={2024}
}



\end{document}